\documentclass[aps,pra,showpacs,twocolumn]{revtex4}
\usepackage{amsmath, amssymb}
\usepackage{graphicx}
\usepackage{subfigure}

\begin{document}

\title{Entanglement in a Raman-driven cascaded system}

\author{C. Di Fidio}

\author{W. Vogel}
\affiliation{Arbeitsgruppe Quantenoptik, Institut f\"ur Physik,
Universit\"at Rostock, D-18051 Rostock, Germany}

\date{July 16, 2008}

\begin{abstract}

The dynamics of a cascaded system that
consists of two atom-cavity subsystems 
is studied by using the quantum trajectory method.
Considering the two atom-cavity subsystems
driven by a Raman interaction, 
analytical solutions are obtained. 
Subsequently,
the entanglement evolution between the two atoms
is studied, and it is shown that the entanglement 
can be stored by switching off the Raman coupling.
By monitoring the radiation field, the entanglement 
between the two atoms can be enhanced.

\end{abstract} 

\pacs{03.67.Bg, 42.50.Pq, 37.30.+i, 42.50.Lc}

\maketitle

\section{Introduction}
\label{introduction}

The concept of entanglement has been of great interest
since the early days of quantum mechanics~\cite{Schroedinger}, 
and it has become of central importance in a variety of 
discussions on the fundamental aspects of the 
theory~\cite{Einstein, Bell}. Nowadays
entanglement is receiving new attention in the context of the
rapidly developing fields of quantum information,
quantum computation and quantum technology; for 
reviews, see~\cite{Nielsen,HarocheRaimond,Kok:135,Chen}.
Entanglement is considered to
be the characteristic feature that allows quantum information
to overcome some of the limitations imposed by classical
information.
Cold trapped atoms 
interacting with quantized light fields are promising
candidates for the realization of quantum computing
and quantum communication protocols~\cite{Monroe:238,Mabuchi:1372}.
The combination of long-lived
atomic states
and light fields can be used in quantum networking for the
distribution and processing of quantum information~\cite{Cirac:3221, Knill:46}.
In the context of entanglement preparation between
atoms at separate nodes, a variety of schemes have 
been proposed, for example,
by measuring the superpositions of light fields 
released from
separate atomic samples, or 
by measuring a probe light field that 
has interacted in a prescribed way with different samples.
Due to the indistinguishability in the measurement,
and conditioned on the results of the measurements,
the atomic system is projected into an entangled 
state~\cite{Bose:5158,Duan:253601,Duan:5643,Duan:413}.
An unconditional preparation of entanglement
has also been analyzed in the case of a cascaded system.
This unconditional preparation has been discussed for
two distantly separated atoms~\cite{Clark:177901,Gu:043813},
as well as for separate atomic ensembles~\cite{Parkins:053602}.
Moreover, the recent achievements in cavity 
QED and in tapped ion techniques have rendered
it possible to experimentally
generate pairs of entangled atoms~\cite{Hagley:1}, to create 
entangled states of several atoms~\cite{Leibfried:639},
and even long-lived entanglement of two macroscopic
ensembles of atoms~\cite{Julsgaard:400}.

In the spirit of these previous achievements, 
in the present contribution
we will consider the quantum trajectory approach for a
cascaded open quantum system~\cite{Carmichael:2273}. 
We study the dynamics of a system that consists of two
atom-cavity sub-systems $A$ and $B$.
The quantum source $A$ emits a photon and the second
quantum subsystem $B$ reacts on the emitted photon.
We will first consider an unconditional
preparation of the entanglement between the two atoms.
Second, the effects of a null-measurement 
conditional preparation is analyzed with respect 
to a photodetector of a given efficiency monitoring 
the field radiated by the cascaded system.

As is clearly discussed in Refs.~\cite{Clark:177901,Gu:043813,Parkins:053602},
the advantage in using a cascaded system is that
the dynamical evolution of the open quantum system itself
creates the entanglement. It is an
unconditional preparation, and it 
is not related to a ``click" or ``no click" at a detector,
where the
measurement projects the atomic system onto the 
desired entangled state. 
In this sense, we can say that it is a dynamical generation 
of entanglement, and not a conditional one.
One could also use 
a detector of given efficiency to monitor
the radiated field,
to prepare the system 
conditioned upon ``no click" 
at the detector.  
This allows us to realize 
a quantum state preparation conditioned upon
the limited knowledge of the observer's imperfect detector. 
In this way one can combine the advantages
of using a cascaded system, with its intrinsic
dynamical generation of entanglement, and a 
conditional preparation with a detector of non-unit
efficiency.
In the case under study 
the entanglement between the
two atoms can only increase due to this conditional
preparation, even for imperfect detection.
In the limiting case of a detector of zero efficiency, 
we return to the case of unconditional preparation,
where only dynamically generated entanglemet is present.
In the system under study we consider a Raman configuration
for driving the atom-cavity interaction. 
By switching off the lasers beams, the Raman
interaction vanishes, 
so that the entanglement generated between the
atoms remains unchanged and can be stored.

The paper is organized as follows.
In Sec.~\ref{section2} the master equation 
describing the dynamics of the cascaded
system is introduced,
and the problem is solved
analytically by using the quantum trajectory method.
In Sec.~\ref{section3} the unconditional 
preparation and storage of the entanglement
between the two atoms is analyzed. 
The conditional preparation of the entanglement
between the two atoms is discussed in Sec.~\ref{section4}.
Finally, some concluding remarks are given in Sec.~\ref{conclusion}.

\section{Cascaded system dynamics}
\label{section2}

\begin{figure}
\includegraphics[width=7.0cm]{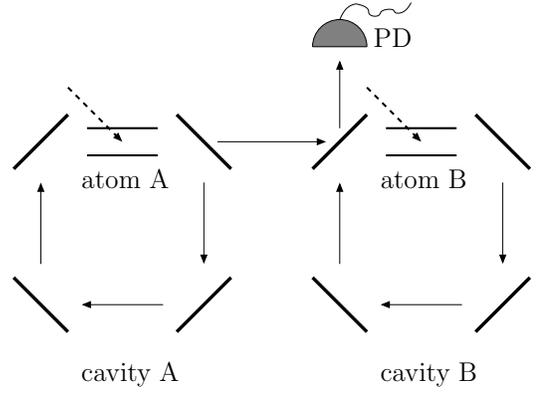}
\caption{The cascaded open system
consisting of two Raman-driven atom-cavity subsystems 
$A$ and $B$. The dashed arrows indicate the laser fields
needed for the Raman coupling.
A photodetector
PD can be used to monitor the radiation field.}
\label{fig:figure_pra_1}
\end{figure}

\begin{figure}
\includegraphics[width=6.0cm]{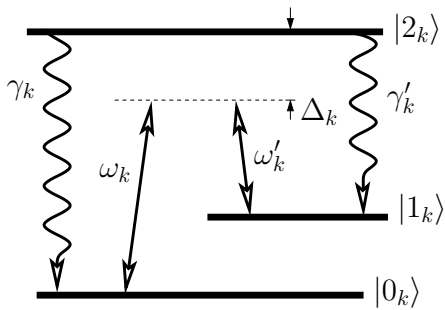}
\caption{Raman-type excitation scheme for the 
atom-cavity subsystem $A$ ($k=a$) and $B$ ($k = b$).
The cavity mode of frequency
$\omega_k$ and the laser of frequency $\omega_k'$ are detuned by $\Delta_k$ from the atomic state $|2_k \rangle$. 
The dipole relaxation rates of these states to the
states $|0_k \rangle$ and $|1_k \rangle$ are $\gamma_k$ and
$\gamma_k'$, respectively.}
\label{fig:figure_pra_2}
\end{figure}

In this section we analyze the dynamics of 
the system under study.
The cascaded open quantum system consists of two
atom-cavity subsystems $A$ and $B$,
where the source subsystem $A$ is cascaded with the 
target subsystem $B$, as sketched in Fig.~\ref{fig:figure_pra_1}.
The cavities have three perfectly reflecting mirrors and one
mirror with transmission coefficient $T \ll 1$.
In the two subsystems $A$ and $B$, denoted by $k=a,b$, respectively,
we consider a three-level
atom coupled to a cavity mode of frequency 
$\omega_k$ via a Raman interaction, 
as indicated in Fig.~\ref{fig:figure_pra_2}. 
This configuration is obtained
by irradiating the atom with a laser beam of frequency
$\omega_k'$ such that $\omega_k-\omega_k'=\omega_{10}^k$, 
where $\omega_{10}^k$ is the transition frequency
between the two atomic energy eigenstates
$|1_k\rangle$ and $|0_k\rangle$.
The laser beam is detuned from the electric dipole transition
$|1_k\rangle \leftrightarrow |2_k\rangle$
by $\Delta_k$,
chosen to enhance the Raman-coupling strength, but 
also to avoid electronic excitations. The Rabi frequency of the laser is denoted by $\Omega_k$ and $g_k$ is the strength of
coupling between the cavity mode and the
$|0_k\rangle \leftrightarrow |2_k\rangle$ transition.
The cavity mode is damped by losses through the partially transmitting cavity mirror. 
In addition to the wanted outcoupling of the field,
the atom can spontaneously emit a photon out the side of the cavity, or a photon can be absorbed or scattered
by the cavity mirrors.

To describe the dynamics of the system we 
will use a master equation formalism, and solve 
it by using the quantum trajectory 
method~\cite{Dalibard:580, Dum:4382, Carmichael1}.
For sufficiently large detuning, $g_k/\Delta_k$ and $\Omega_k/\Delta_k \ll 1$, 
the  excited state $|2_k\rangle$ will not
become significantly populated and 
can be adiabatically eliminated.
This leads,
treating the dissipation due to the cavity losses
in a standard way~\cite{Haake, Louisell, Davies}, to the
following master equation
for the reduced density 
operator $\hat \rho(t)$ of the system:
\begin{eqnarray}
\frac{d\hat\rho(t)}{dt} \!\!\!&=&\!\!\! \frac{1}{i\hbar} \! \left[\hat H,\hat\rho(t)\right] \!+\! \sum_{i = 1}^7
\left[ \hat J_i \hat\rho(t) \hat J_i^\dagger 
- \frac{1}{2} \hat J_i^\dagger \hat J_i \hat\rho(t) \right. \nonumber \\
&-& \left. \frac{1}{2} \hat\rho(t) \hat J_i^\dagger \hat J_i \right] .
\label{eq:master_1}
\end{eqnarray}
The Hamiltonian is given by
\begin{equation}
\hat H = \hat H_A + \hat H_B + i \hbar \frac{\sqrt{\kappa_a \kappa_b}}{2}
\left( e^{-i \phi} \hat b \hat a^\dagger - e^{i \phi} \hat b^\dagger \hat
a \right) \, ,
\label{eq:hamiltonian}
\end{equation}
where $\hat H_A$ and $\hat H_B$ describe the atom-cavity interaction in the two subsystems $A$ and $B$, respectively.
In the rotating-wave approximation they are given
by~\cite{Difidio:105}
\begin{equation}
\hat H_A = \hbar \bar g_a  \left( \hat{a} \hat A_{10} 
+ \hat a^\dagger \hat A_{01}\right) + \hbar \Delta_a' {\hat A}_{11}
+ \hbar \bar \Delta_a \hat a^\dagger \hat{a} {\hat A}_{00}  \, ,
\label{eq:JC_hamiltonian_A}
\end{equation}
and
\begin{equation}
\hat H_B = \hbar \bar g_b \left( \hat{b} \hat B_{10} 
+ \hat b^\dagger \hat B_{01}\right) + \hbar \Delta_b' {\hat B}_{11}
+ \hbar \bar \Delta_b \hat b^\dagger \hat{b} {\hat B}_{00}   \, .
\label{eq:JC_hamiltonian_B}
\end{equation}
The third term in Eq.~(\ref{eq:hamiltonian}) describes the
coupling between the two
cavities~\cite{Carmichael:2273, Gardiner:2269}. 
In these expressions, $\hat a$ and 
$\hat a^\dagger$ 
are annihilation and creation operators
for the cavity field $A$,
and similarly $\hat b$ and 
$\hat b^\dagger$ for the cavity field $B$. 
We have also defined $\hat A_{ij} = |i_a\rangle \langle j_a|$ ($i,j = 0, 1$),
and $\hat B_{ij} = |i_b\rangle \langle j_b|$ ($i,j = 0, 1$).
In addition, $\bar g_k \!=\! -g_k\Omega_k/ \Delta_k$
is the effective 
atom-cavity coupling constant
and $\Delta_k' \!=\! -\Omega_k^2 / \Delta_k$,
$\bar \Delta_k \!=\! -g_k^2 / \Delta_k$ are the two
Stark shift terms.
Moreover, $\kappa_a$, $\kappa_b$ are the cavity bandwidths
and the phase $\phi$ is related to the phase change upon reflection from the source output mirror,
and/or to the retardation of the source due
to the spatial separation between the source and the target,
cf.~\cite{Carmichael2}.

The jump operators $\hat J_i$ in Eq.~(\ref{eq:master_1}) are defined by
\begin{equation}
\hat J_1 = \sqrt{\kappa_a} \hat a + \sqrt{\kappa_b} e^{-i\phi}
\hat b \, ,
\label{jump_1}
\end{equation}
which describes a photon emission by the cavities; 
\begin{equation}
\hat J_2 = \sqrt{\kappa_a'} \hat a \, , ~~~~~~~
\hat J_3 = \sqrt{\kappa_b'} \hat b \, ,
\label{jumps_2_3}
\end{equation}
are associated with a photon absorption or scattering
by the cavity mirrors; 
\begin{eqnarray}
\hat J_4 &=& \sqrt{\Gamma_a} \! \left( \! \Omega_a \hat A_{01}
\!+\!
g_a \hat{a}  \hat A_{00} \! \right)  , \nonumber \\
\hat J_5 &=& \sqrt{\Gamma_a'} \! \left(\! \Omega_a \hat A_{11}
\!+\!
g_a \hat{a}  \hat A_{10} \! \right) \! ,
\label{jumps_4_5}
\end{eqnarray}
and
\begin{eqnarray}
\hat J_6 &=& \sqrt{\Gamma_b} \! \left( \! \Omega_b \hat B_{01}
\!+\!
g_b \hat{b}  \hat B_{00} \! \right) ,  \nonumber \\
\hat J_7 &=& \sqrt{\Gamma_b'} \! \left( \! \Omega_b \hat B_{11}
\!+\!
g_b \hat{b}  \hat B_{10} \! \right) \! ,
\label{jumps_6_7}
\end{eqnarray}
are related to a photon spontaneously emitted by the atoms.
Here $\kappa_k'$ is
the cavities mirrors' absorption and scattering rate.
Moreover,
$\Gamma_k \!=\! \gamma_k /\Delta_k^2$ and
$\Gamma_k ' \!=\! \gamma_k'/ \Delta_k^2$,
where $\gamma_k$ and $\gamma_k'$ are the dipole relaxation rates of the atomic state $|2_k\rangle$ to the states $|0_k\rangle$
and $|1_k\rangle$,  respectively. These relaxation rates
are considered to be small in comparison with the detuning. 
Note that the operator $\hat J_1$ contains 
the superposition of the two fields radiated by the
two cavities, due to the fact that radiated photons 
cannot be associated
with photon emission from either $A$ or $B$ separately~\cite{Carmichael:2273}.

In the following we will identify, for notational convenience,
the state $|a\rangle$ with the state $|1,0,0,0\rangle$, 
which denotes the atom $A$ in
the state $|1_a\rangle$, the cavity $A$ in the vacuum
state, the atom $B$ in the state $|0_b\rangle$ , and
the cavity $B$ in the vacuum state.
In the state $|b\rangle \equiv |0,1,0,0\rangle$
the atom $A$ is in
the state $|0_a\rangle$, and the cavity $A$ is in the 
one-photon Fock state.
Similarly, we define $|c\rangle \equiv |0,0,1,0\rangle$,
$|d\rangle \equiv |0,0,0,1\rangle$, and
$|e\rangle \equiv |0,0,0,0\rangle$.
The state $|a\rangle$ will be considered as the initial
state of the system. It follows that the Hilbert space that describes the cascaded system under study is, in our 
model, spanned by the five state vectors $|a\rangle$, $|b\rangle$, $|c\rangle$, $|d\rangle$, and $|e\rangle$.

To evaluate the time evolution of the system 
we use a quantum trajectory
approach~\cite{Dalibard:580, Dum:4382, Carmichael1}. 
Note that
the probability for a jump 
$\hat J_i$ to occur in the time 
interval $[t, t + dt )$ is given by $p_{\rm i}(t) \!=\! 
\langle \hat J_i^\dagger \hat J_i  \rangle_t \, d t$.
This implies that the total probability
for a jump due to a spontaneous emission
in the time interval $[t, t + dt )$ is given,
using Eqs.~(\ref{jumps_4_5}), (\ref{jumps_6_7}), (\ref{eq:JC_hamiltonian_A}), 
and (\ref{eq:JC_hamiltonian_B}), by
\begin{eqnarray}
\sum_{i=4}^7\langle \hat J_i^\dagger \hat J_i  \rangle_t  d t
= - \frac{\gamma_a + \gamma_a'}{\hbar \Delta_a}\langle \hat H_A \rangle_t d t - \frac{\gamma_b + \gamma_b'}{\hbar
\Delta_b}
\langle\hat H_B\rangle_t d t. 
\end{eqnarray}
This relation shows that in a time interval 
$\sim \Delta/(g \Omega)$ the probability to have
a jump due to spontaneous emissions 
is $\sim (\gamma_a \!+ \gamma_a' \!+ \gamma_b \!+ \gamma_b')/\Delta$, cf.~\cite{Wineland:2977}.
For a large detuning this probability is small. 
If one is interested to follow the
dynamical evolution of the system for several 
Rabi oscillations, in general the effects due to spontaneous 
emissions cannot be neglected~\cite{Difidio:031802}.
In the present contribution, the Raman dynamics 
would be actually used only for a few Rabi
oscillations and we may neglect the terms in
the master equation related to spontaneous emissions.

Let us now consider the system prepared at time
$t_0 = 0$ in the state $|a\rangle$. To determine 
the state vector of the system at a later time $t$, 
provided that no jump has occurred between time $t_0$ and $t$,  
we have to solve the nonunitary Schr\"{o}dinger
equation
\begin{equation}
i \hbar \frac{d}{dt} 
| \bar{\psi}_{\rm no} (t) \rangle  =\hat{H^{'}} \, | \bar{\psi}_{\rm no} (t) \rangle \, ,
\label{eq:schr_non_unitary} 
\end{equation}
where $\hat{H^{'}}$ is the non-Hermitian Hamiltonian
given by
\begin{eqnarray}
\hspace*{-0.2cm}
\hat{H^{'}} \!\!&=&\!\! \hat H
- \frac{i \hbar}{2} \sum_i^3 \hat J_i^\dagger
\hat J_i = \hat H_A + \hat H_B - i \hbar \Big(
\frac{{\cal{K}}_a}{2}
\hat a^\dagger \hat a  \nonumber \\
\!\!&+&\!\! \frac{{\cal{K}}_b}{2} \hat b^\dagger \hat b 
+ 
\sqrt{\kappa_a \kappa_b} e^{i \phi} \hat b^\dagger \hat a \Big) ,
\label{eq:n_H_Hamiltonian}
\end{eqnarray}
where we have defined
\begin{equation}
{{\cal{K}}_a} = \kappa_a + \kappa_a' \, , ~~~~
{{\cal{K}}_b} = \kappa_b + \kappa_b' \, .
\label{eq:kappa}
\end{equation}
If no jump has occurred between time $t_0$ and $t$, the system evolves via Eq.~(\ref{eq:schr_non_unitary}) into 
the unnormalized state
\begin{equation}
| \bar{\psi}_{\rm no} (t) \rangle = \alpha(t) |a\rangle + \beta(t) |b \rangle + \gamma(t) |c\rangle + \delta(t) |d \rangle \, .
\label{eq:psi_n}
\end{equation}
In this case the conditioned density operator for
the atom-cavity system is given by 
\begin{eqnarray}
\hat \rho_{\rm no} (t) &=& \frac{ | \bar{\psi}_{\rm no} (t) \rangle \langle
\bar{\psi}_{\rm no} (t) |}{\langle \bar{\psi}_{\rm no} (t) |\bar{\psi}_{\rm no} (t) \rangle}
\, .
\label{eq:rho_no}
\end{eqnarray}
Here we have used the word conditioned to stress the fact that 
this is the density operator at time $t$, conditioned on the
fact that no jump has occurred between time $t_0$ and $t$.

The evolution governed by the nonunitary Schr\"odinger equation~(\ref{eq:schr_non_unitary})
is randomly interrupted by one of the three
kinds of jumps $\hat J_i$,
cf. Eqs.~(\ref{jump_1}) and (\ref{jumps_2_3}).
If a jump has occurred at time
$t_{\rm J}$, $t_{\rm J} \in (t_0, t]$, the
state vector is collapsed in the state $|e \rangle$ due to the action of one of the jump operators,
\begin{equation}
\hspace*{-0.5cm} \hat J_{i} \, | \bar{\psi}_{\rm no} (t_{\rm J}) \rangle  \rightarrow |e \rangle   ~~(i = 1, 2, 3). \label{eq:jump_op_i} 
\end{equation}
In the problem under study we may have only one jump. Once the system collapses into the state $|e\rangle$, the nonunitary Schr\"odinger equation~(\ref{eq:schr_non_unitary}) 
lets it remain unchanged.
In this case the conditioned density operator at time $t$ is given by
\begin{equation}
\hat \rho_{\rm yes} (t) = | e \rangle \langle e | \, ,
\label{eq:rho_yes}
\end{equation}
where we indicate with ``yes" the fact that a jump has occurred.

In the quantum trajectory method, the density operator $\hat \rho (t)$ is obtained by performing an ensemble
average over the different conditioned density operators at
time $t$. In the present case, starting at time $t_0$ with the density operator $\hat \rho_0 = | a \rangle \langle a |$,
the ensemble average is performed over the two 
possible realizations (histories) ``yes" and ``no",
yielding the statistical mixture
\begin{equation}
\hat \rho (t) =  p_{\rm no}(t) \hat \rho_{\rm no} (t) +
  p_{\rm yes}(t) \hat \rho_{\rm yes} (t) \, .
\label{eq:rho_t}
\end{equation}
Here $p_{\rm no}(t)$ and $p_{\rm yes}(t)$ are the probability that between the initial time $t_0$ and time $t$ no jump and
one jump has occurred, respectively. Of course, 
$p_{\rm no}(t) + p_{\rm yes}(t) = 1$.

To evaluate $p_{\rm no}(t)$, we use the method of the delay function~\cite{Dum:4382}. This yields the
probability $p_{\rm no}(t)$ as the square of 
the norm of the unnormalized state vector:
\begin{eqnarray}
p_{\rm no}(t) 
&=& \parallel | \bar{\psi}_{\rm no} (t) \rangle \! \parallel^2
= \langle \bar{\psi}_{\rm no} (t) |\bar{\psi}_{\rm no} (t) \rangle \nonumber \\ 
&=& |\alpha(t)|^2  + |\beta(t)|^2 + |\gamma(t)|^2  
+ |\delta(t)|^2\, .
\label{eq:p_no}
\end{eqnarray}
From Eqs.~(\ref{eq:rho_t}) and (\ref{eq:p_no}) one obtains 
\begin{eqnarray}
\hat \rho (t) =  
| \bar{\psi}_{\rm no} (t) \rangle \langle
\bar{\psi}_{\rm no} (t) |
+ |\epsilon(t)|^2 |e \rangle \langle e| \, ,
\label{eq:rho_t_1}
\end{eqnarray}
where we have defined 
\begin{equation}
|\epsilon(t)|^2 = p_{\rm yes}(t) = 1 - p_{\rm no}(t) \, .
\label{eq:p_yes}
\end{equation}
The quantities $|\alpha(t)|^2$, $|\beta(t)|^2$, $|\gamma(t)|^2$, $|\delta(t)|^2$, and $|\epsilon(t)|^2$ 
represent the probabilities that at time $t$ the system can be found either in $|a\rangle$,
$|b\rangle$, $|c\rangle$, $|d\rangle$, and $|e\rangle$,
respectively.

To determine $\alpha(t)$, $\beta(t)$,
$\gamma(t)$, and $\delta(t)$, we have to solve
the nonunitary Schr\"odinger 
equation~(\ref{eq:schr_non_unitary}) together
with~(\ref{eq:n_H_Hamiltonian}). 
This leads to the inhomogeneous system 
of differential equations,
\begin{equation}
\left\{ 
\begin{array}{llll}
\dot \alpha(t) 
= -i \Delta_a' \alpha (t) - i \bar g_a \beta(t) \, , \\
\dot \beta(t) = - i \bar g_a \alpha(t) - ({\cal{K}}_a/2 + i
\bar \Delta_a) \beta(t) 
\, ,
\\
\dot \gamma(t) = -i \Delta_b' \gamma (t) - i \bar g_b \delta(t) \, ,  \\
\dot \delta(t) = - i \bar g_b \gamma(t) - ({\cal{K}}_b/2 + i 
\bar \Delta_b) \delta(t) 
- \sqrt{\kappa_a\kappa_b} e^{i \phi} \beta (t) \, .
\end{array}
\right.
\label{eq:sys_diff_eq_1}
\end{equation}
The differential equations for $\alpha(t)$ and
$\beta(t)$ can be solved independently from
those for $\gamma(t)$ and $\delta(t)$.
For the initial conditions $\alpha(0) \!=\!1$ and 
$\beta(0) \!=\! 0$, i.e. at time $t=0$ the atom $A$ is in 
the state $|1_a\rangle$ and the cavity $A$ in the vacuum state,
and defining
\begin{equation}
\Lambda_a \!\equiv\! \sqrt{\frac{\left({\cal{K}}_a \!+\! 2i \bar \Delta_a \right)^2}{4} \!-\! 4 \bar g_a^2 \!-\! i \left(
{\cal{K}}_a \!+\! 2i\bar \Delta_a \right) \Delta_a' \!-\! {\Delta_a'}^2 } \, , 
\end{equation}
we can write the solutions
for $\alpha(t)$ and $\beta(t)$, similarly as done 
in~\cite{Difidio}, as
\begin{eqnarray}
&& \hspace*{-0.5cm} {\alpha} (t) \!=\!  \left[\frac{
({\cal{K}}_a + 2i\bar \Delta_a)/2 - i \Delta_a'}{\Lambda_a} \sinh \left(\frac{\Lambda_a t}{2}\right)  \right.
\nonumber \\
&& ~~~~~~~~ + \left. \cosh \left( \frac{\Lambda_a t}{2}\right) \right] \! e^{-[({\cal{K}}_a +2i\bar \Delta_a)/4 
+ i \Delta_a'/2]t} , \nonumber \\
&& \hspace*{-0.5cm} {\beta} (t) \!=\! - \frac{2i\bar g_a}{\Lambda_a} \sinh \left(\frac{\Lambda_a t}{2}\right) e^{-[({\cal{K}}_a + 
2i\bar \Delta_a)/4 + i\Delta_a'/2]t} .
\label{eq:diff_eq_sol_general_a_b}
\end{eqnarray}
Inserting now in the inhomogeneous pair
of differential equations for $\gamma(t)$ and $\delta(t)$
the solution obtained for $\beta(t)$,
we can determine the solutions for
$\gamma(t)$ and $\delta(t)$ with the method of the
fundamental matrix. For the initial conditions
$\gamma(0) \!=\!0$ and $\delta(0) \!=\! 0$, 
i.e. at time $t=0$ the atom $B$ is 
in the state $|0_b\rangle$ and the cavity $B$ in the vacuum state,
and defining
\begin{equation}
\Lambda_b \!\equiv\! \sqrt{\frac{\left({\cal{K}}_b \!+\! 2i\bar \Delta_b \right)^2}{4} \!-\! 4 \bar g_b^2 \!-\! i \left(
{\cal{K}}_b \!+\! 2i\bar \Delta_b \right) \Delta_b' \!-\! {\Delta_b'}^2 } \, , 
\end{equation}
we get
\begin{eqnarray}
\hspace*{-0.18cm} {\gamma} (t) \!\!&=&\!\! \bar g_b \left\{
f_{+}(t) \! \left[g_{-}(t) \!+\! h_{+}(t)  \right] 
\!-\! f_{-}(t) \!
\left[g_{+}(t) \!+\! h_{-}(t)  \right] \right\} ,
\nonumber \\
\hspace*{-0.18cm} {\delta} (t) \!\!&=&\!\! 
i \! \left[\frac{{\cal{K}}_b \!+\! 2 i \bar \Delta_b}{4} \!-\! 
i \frac{\Delta_b'}{2} \!+\! \frac{\Lambda_b}{2}\right] \! f_{-}(t) 
\! \left[g_{+}(t) \!+\! h_{-}(t)  \right] \nonumber \\
\!\!&-&\!\! i \! \left[\frac{{\cal{K}}_b \!+\! 2 i \bar \Delta_b}{4} \!-\!
i \frac{\Delta_b'}{2} \!-\! \frac{\Lambda_b}{2}\right] \! f_{+}(t) 
\! \left[g_{-}(t) \!+\! h_{+}(t)  \right] \!.
\label{eq:diff_eq_sol_general_g_d}
\end{eqnarray}
Here we have defined, for notational convenience,
\begin{equation}
f_\pm(t) = \frac{\bar g_a \sqrt{\kappa_a\kappa_b} e^{i \phi} }{\Lambda_a \Lambda_b} e^{[-({\cal{K}}_b + 2 i \bar \Delta_b)/4 - i \Delta_b'/2 \pm \Lambda_b/2]t} \, ,
\label{eq:fpm}
\end{equation}
\begin{equation}
g_\pm(t) = \frac{e^{[(\Lambda_a\pm \Lambda_b)/2 - \Upsilon - i \Theta]t}-1}{(\Lambda_a\pm \Lambda_b)/2 - \Upsilon - i \Theta} \, ,
\end{equation}
and
\begin{equation}
h_\pm(t) = \frac{e^{-[(\Lambda_a\pm \Lambda_b)/2 + \Upsilon + i \Theta]t}-1}{(\Lambda_a\pm \Lambda_b)/2 + \Upsilon + i \Theta}  \, ,
\end{equation}
where $\Upsilon \!=\! ({\cal{K}}_a - {\cal{K}}_b + 2i\bar \Delta_a \!-\! 2i\bar \Delta_b)/4$, and $\Theta \!=\! (\Delta_a' - \Delta_b')/2$. 
In the case of equal parameters for the
two subsystems $A$ and $B$, the solutions (\ref{eq:diff_eq_sol_general_g_d}) simplify as 
\begin{eqnarray}
{\gamma} (t) \!\!&=&\!\! \frac{\kappa \bar g^2 e^{i \phi}}{\Lambda^3}
\left[e^{-\Lambda t} \!+\! \Lambda t \!-\! 1 \right] 
e^{[-({\cal{K}} +2 i \bar \Delta)/4 - i \Delta'/2 + \Lambda/2]t} 
\nonumber \\
\!\!&-&\!\! \frac{\kappa \bar g^2 e^{i \phi}}{\Lambda^3}
\left[e^{\Lambda t} \!-\! \Lambda t \!-\! 1 \right]
e^{[-({\cal{K}} + 2 i \bar \Delta)/4 - i \Delta'/2 - \Lambda/2]t} \, ,
\nonumber \\
{\delta} (t) \!\!&=&\!\! \frac{i\kappa \bar g e^{i\phi}}{\Lambda^3}
\left[\frac{{\cal{K}} \!+\! 2 i \bar \Delta}{4} \!-\! i \frac{\Delta'}{2}
\!+\! \frac{\Lambda}{2}\right]\left[e^{\Lambda t}
- \Lambda t - 1 \right] \nonumber \\
\!\!&\times&\!\! 
e^{[-({\cal{K}} + 2 i \bar \Delta)/4 - i \Delta'/2 - \Lambda/2]t} \!
- \frac{i\kappa \bar g e^{i \phi}}{\Lambda^3}\left[\frac{{\cal{K}}\!+\!
2 i \bar \Delta}{4} \!-\! i \frac{\Delta'}{2} \right. \nonumber \\
\!\!&-&\!\! \left. \frac{\Lambda}{2}\right] \left[e^{-\Lambda t} \!+\! \Lambda t \!-\! 1 \right]  e^{[-({\cal{K}} + 2 i \bar \Delta)/4 - i \Delta'/2 + \Lambda/2]t} .
\label{eq:diff_eq_sol_equal_values}
\end{eqnarray}
where we have used
$\lim_{x \to 0} \{[\exp(\pm x t) - 1]/x\} = \pm t$, and
defined
$\kappa \!=\! \kappa_a \!=\! \kappa_b$,
${\cal{K}} \!=\! {\cal{K}}_a \!=\! {\cal{K}}_b$,
$\bar g \!=\! \bar g_a \!=\! \bar g_b$,
$\bar \Delta \!=\! \bar \Delta_a \!=\! \bar \Delta_b$, 
$\Delta' \!=\! \Delta_a' \!=\! \Delta_b'$,
and $\Lambda \!=\! \Lambda_a \!=\! \Lambda_b$.

\begin{figure}
\includegraphics[width=8cm]{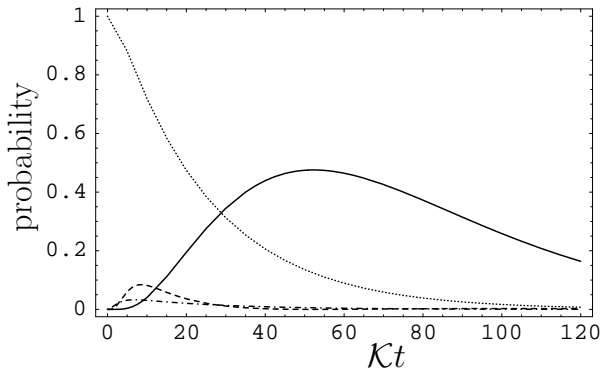}
\caption{The probabilities $|\alpha(t)|^2$ (dotted line), $|\beta(t)|^2$ (dot-dashed line), $|\gamma(t)|^2$ (solid line), 
and $|\delta(t)|^2$ (dashed line) are shown for 
the case of equal parameters for the
two subsystems $A$ and $B$, where $g/{\cal{K}} = 10$,
$\Omega/{\cal{K}} = 10$, $\Delta/{\cal{K}} = 1000$,
$\kappa/{\cal{K}} = 0.9$, and $2(\gamma + \gamma')/\Delta \ll 1$.}
\label{fig:figure_pra_3}
\end{figure}

Using the solutions given by Eqs.~(\ref{eq:diff_eq_sol_general_a_b}) and (\ref{eq:diff_eq_sol_general_g_d}), or (\ref{eq:diff_eq_sol_equal_values}), one can plot
the functions
$|\alpha(t)|^2$, $|\beta(t)|^2$, $|\gamma(t)|^2$, 
and $|\delta(t)|^2$, i.e.
the occupation probabilities of the states $|a\rangle$,
$|b\rangle$, $|c\rangle$, and $|d\rangle$, respectively.
In Fig.~\ref{fig:figure_pra_3} we show these
probabilities
for the case of equal parameters for the
two subsystems $A$ and $B$, with 
$g/{\cal{K}} = 10$, $\Omega/{\cal{K}} = 10$, $\Delta/{\cal{K}} = 1000$, and $\kappa/{\cal{K}} = 0.9$, i.e. the absorption or scattering by the cavity mirrors is $10\%$ 
of the total cavity decay.
Note that the phase factor $e^{i\phi}$ does
not play any role in the functions considered here.
From the figure one can see how the dynamical evolution of the source subsystem $A$ drives the target subsystem $B$.
Of course, for ${\cal{K}}t \gg 1$, these propabilities 
are all tending to zero, due to the fact that, sooner or
later, a photon is absorbed or scattered by the cavities
mirrors, or is emitted into the radiated field, so that
the state vector of the system is projected onto 
the state $|e\rangle$.

\section{Unconditional preparation of entanglement}
\label{section3}

In this section we study the dynamical generation
of the entanglemet between the two subsystem $A$ and $B$.
In particular
the entanglement evolution 
will be 
analyzed by means of the concurrence. 
We will also see that
the entanglement generated between the two atoms
can be stored by switching off the Raman interaction.

\subsection{Entanglement evolution}

For the situation under study,
the two atoms constitute a pair of qubits.
An appropriate measure of the entanglement
for a two qubits system, often considered in the context of
quantum information theory, is the concurrence~\cite{Wootters:2245}.
Given the density matrix $\rho$ for such a system,
the concurrence is defined as
\begin{equation}
C(\rho) = \max \left\{ 0, \sqrt{\lambda_1} -
\sqrt{\lambda_2}- \sqrt{\lambda_3}- \sqrt{\lambda_4} 
\right\} \, ,
\label{eq:conc}
\end{equation}
where $\lambda_1 \geq  \lambda_2 \geq \lambda_3
\geq \lambda_4$ are the eigenvalues of the matrix
$\tilde \rho = \rho (\sigma_y \otimes \sigma_y)
\rho^{*} (\sigma_y \otimes \sigma_y)$.
Here $\sigma_y$ 
is the Pauli
spin matrix and complex conjugation is denoted by an 
asterisk.
The concurrence varies in the range $[0,1]$,
where the values $0$ and $1$ represent separable states
and maximally entangled states, respectively.

\begin{figure}
\includegraphics[width=8.0cm]{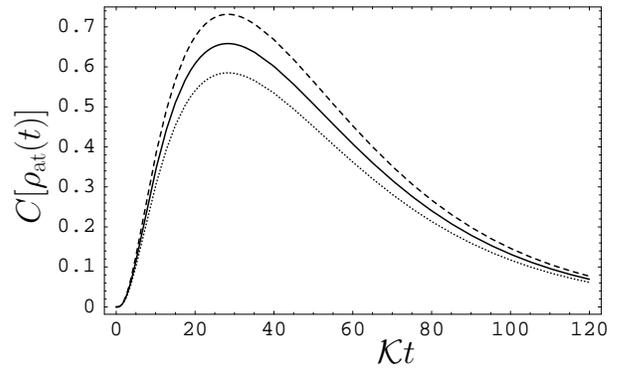}
\caption{The concurrence $C[\rho_{\rm at}(t)]$ between
the two atoms is shown for 
the case of equal parameters for the
two subsystems $A$ and $B$, where 
$g/{\cal{K}} = 10$, $\Omega/{\cal{K}} = 10$, $\Delta/{\cal{K}} = 1000$, 
$2(\gamma + \gamma')/\Delta \ll 1$,
and $\kappa/{\cal{K}} = 1$ (dashed line),
$\kappa/{\cal{K}} = 0.9$ (solid line),
$\kappa/{\cal{K}} = 0.8$ (dotted line).}
\label{fig:figure_pra_4}
\end{figure} 
To derive an expression for the  
concurrence between the two atoms,
let us consider
the density operator that describes the system.
It is obtained from
the density operator $\hat \rho (t)$, Eq.~(\ref{eq:rho_t_1}),
by tracing over the intracavity field states for 
the two subsystems, $\hat \rho_{\rm at}(t) \!=\! {\rm Tr}_{\rm cav} 
\left[ \hat \rho (t) \right]$,
and is given by
\begin{eqnarray}
\hat \rho_{\rm at}(t) \!\!&=&\!\!  
|\alpha(t)|^2 |1,0 \rangle \langle 1,0| +
|\gamma(t)|^2 |0,1 \rangle \langle 0,1| \nonumber \\
\!\!&+&\!\! \alpha(t)\gamma^{*}(t)|1,0 \rangle \langle 0,1|
+\alpha^{*}(t)\gamma(t)|0,1 \rangle \langle 1,0|
\nonumber \\
\!\!&+&\!\! \left\{ 1 - \left[|\alpha(t)|^2 + |\gamma(t)|^2   
\right]  \right\} |0,0 \rangle \langle 0,0| \, .
\label{eq:rho_atoms}
\end{eqnarray}
Considering the $4\times4$ density matrix 
$\rho_{\rm at}(t)$, related to the density operator
in Eq.~(\ref{eq:rho_atoms}) in the two-atom basis 
$\{ |0,0 \rangle, |0,1 \rangle, |1,0 \rangle,
|1,1 \rangle \}$, it is easy to show that the
concurrence $C[\rho_{\rm at}(t)]$ is, using 
Eq.~(\ref{eq:conc}), given by
\begin{equation}
C[\rho_{\rm at}(t)] = 2 \left| \alpha(t) \right| \left| 
\gamma(t)  \right| \, .
\label{eq:conc_atoms}
\end{equation}
To analyze the time dependence of this concurrence,
let us consider the case of equal parameters for the
two subsystems $A$ and $B$. Inserting
the analytical solutions 
(\ref{eq:diff_eq_sol_general_a_b}) and (\ref{eq:diff_eq_sol_equal_values})
into Eq.~(\ref{eq:conc_atoms}), we show
in Fig.~\ref{fig:figure_pra_4}
the function $C[\rho_{\rm at}(t)]$
for the parameters
$g/{\cal{K}} = 10$, $\Omega/{\cal{K}} = 10$, and $\Delta/{\cal{K}} = 1000$, for
different values of $\kappa/{\cal{K}}$.
Since the concurrence contains only
absolute values, the phase 
factor $e^{i \phi}$ does not play any role here.
From this figure one can clearly see that the
initially disentangled atoms become
entangled. In particular,
a maximum value for $C[\rho_{\rm at}(t)]$ is
found for $\bar t \simeq 28.32/{\cal{K}}$, where,
for the shown cases,
$C[\rho_{\rm at}(\bar t)] \simeq 0.73,~0.66$, and $0.59$.
Note that the effects due to the absorption
or scattering by the cavity mirrors are not negligible.
For example, the relative variation of
the concurrence is approximately $10\%$ between
the case $\kappa/{\cal{K}}=1$
(no absorption or scattering) and $\kappa/{\cal{K}}=0.9$,
considering the peak at $\bar t$.
Of course, for ${\cal{K}} t \gg 1$, the two atoms become 
again disentangled due to the emission
of the photon in one of the three decay channels.
This is in agreement with the fact that 
the release of a photon into the environment destroys 
any entanglement, projecting the two-atom subsystem 
into the separable state $|0,0\rangle$.
The inclusion of the very rare spontaneous emissions would
only speed up somewhat the decay of the entanglement. 

Finally, we note that the concurrence between
the two intracavity fields can be obtained as well.
Let us consider 
the density operator that describes the system
of the two intracavity fields $A$ and $B$,
obtained from the density operator $\hat \rho (t)$, cf. Eq.~(\ref{eq:rho_t_1}), by tracing over the atomic states
of the two subsystems,
$\hat \rho_{\rm cav}(t) \!=\! {\rm Tr}_{\rm at} 
\left[ \hat \rho (t) \right]$.
It is given by
\begin{eqnarray}
\hat \rho_{\rm cav}(t) \!\!&=&\!\! 
|\beta(t)|^2 |1,0 \rangle \langle 1,0| +
|\delta(t)|^2 |0,1 \rangle \langle 0,1| \nonumber \\
\!\!&+&\!\! \beta(t)\delta^{*}(t)|1,0 \rangle \langle 0,1|
+\beta^{*}(t)\delta(t)|0,1 \rangle \langle 1,0|
\nonumber \\
\!\!&+&\!\! \left\{ 1 - \left[|\beta(t)|^2 + |\delta(t)|^2   
\right]  \right\} |0,0 \rangle \langle 0,0| \, .
\label{eq:rho_cavities}
\end{eqnarray}
Considering now the $4\times4$ density matrix 
$\rho_{\rm cav}(t)$ 
in the two intracavity-fields Fock basis 
$\{ |0,0 \rangle, |0,1 \rangle, |1,0 \rangle,
|1,1 \rangle \}$, 
the concurrence $C[\rho_{\rm cav}(t)]$ is given by
\begin{equation}
C[\rho_{\rm cav}(t)] = 2 \left| \beta(t) \right| \left| 
\delta(t)  \right| \, .
\label{eq:conc_cavities}
\end{equation}
Note that the concurrence for the intracavity fields
is of the same form as the one for the two atoms, cf. Eq.~(\ref{eq:conc_atoms}), when replacing
$\alpha(t)$ and $\gamma(t)$ with $\beta(t)$ and $\delta(t)$.
 
\subsection{Storage of entanglement}

We analyze now the possibility to store the
entanglement between the two atoms.
As in the previous subsection,
let us indicate with $\bar t$ the time
when $C[\rho_{\rm at}(t)]$ reaches its maximum value.
We consider the case when
at time $t = \bar t$ we switch off the two
lasers, i.e. $\Omega_a \!=\! \Omega_b \!=\! 0$,
so that the Raman coupling vanishes.
For $t > \bar t$
the inhomogeneous system of differential equations
becomes
\begin{equation}
\left\{ 
\begin{array}{llll}
\dot \alpha(t) 
= 0 \, , \\
\dot \beta(t) = - ({\cal{K}}_a/2 + i \bar \Delta_a) \beta(t) 
\, ,
\\
\dot \gamma(t) = 0 \, ,  \\
\dot \delta(t) =  - ({\cal{K}}_b/2 + i \bar \Delta_b) \delta(t) 
- \sqrt{\kappa_a\kappa_b} e^{i \phi} \beta (t) \, .
\end{array}
\right.
\label{eq:sys_diff_eq_2}
\end{equation}
It is immediate to write the solutions for $\alpha(t)$, $ \gamma(t)$, and $\beta(t)$ for $t \geq \bar t$ as
\begin{eqnarray}
\alpha(t) \!&=&\! \alpha(\bar{t} )\,,~~~ 
\beta(t) = \beta(\bar t) e^{- ({\cal{K}}_a/2 + i \bar \Delta_a)(t-\bar t)} \,, \nonumber \\
\gamma(t) \!&=&\! \gamma(\bar t)\,.
\label{eq:sol_tbar1}
\end{eqnarray}
Using the solution for $\beta(t)$ in the differential equation for $\delta(t)$, one gets, for $t \geq \bar t$,
\begin{eqnarray}
\delta(t) &=&  e^{- ({\cal{K}}_b/2 + i \bar \Delta_b)(t-\bar t)} \Big[ \delta(\bar t) - \sqrt{\kappa_a\kappa_b} 
e^{i\phi} \beta(\bar t)  \nonumber \\
&\times&  \frac{e^{[- ({\cal{K}}_a - {\cal{K}}_b)/2 - i (\bar \Delta_a - \bar \Delta_b) ](t-\bar t)} -1}{- ({\cal{K}}_a - {\cal{K}}_b)/2 - i (\bar \Delta_a - \bar \Delta_b)} \,  \Big] .
\label{eq:delta_tbar}
\end{eqnarray}
For equal parameters in the two
subsystems Eq.~(\ref{eq:delta_tbar}) 
reads as
\begin{equation}
\delta(t) =  e^{- ({\cal{K}}/2 + i \bar \Delta)(t-\bar t)} \Big[ \delta(\bar t) - \kappa 
e^{i\phi} \beta(\bar t) (t-\bar t) \Big] \, ,
\label{eq:delta_tbar_equal}
\end{equation}
where we have used
$\lim_{x \to 0} \{\exp[x (t\!-\!\bar t)] \!-\! 1\}/x 
\!=\!(t\!-\!\bar t)$.

\noindent
Using these solutions it is clear that, for $t \geq \bar t$,
the concurrence
between the two atoms remains constant and 
its value is
\begin{equation}
C[\rho_{\rm at}(t)] = C[\rho_{\rm at}(\bar t)] =2 \left| \alpha(\bar t) \right| \left| 
\gamma(\bar t)  \right| \, .
\label{eq:conc_atoms_tbar}
\end{equation}
This expression shows that the entanglement between the 
two atoms can be stored even for times with ${\cal{K}}t \gg 1$. 
This situation could be realized, for example, 
by using a hyperfine transition in $^9{\rm Be}^+$ ions,
whose coherence time between the two internal
levels was reported to be several
minutes~\cite{Wineland:259}. 
The experimental setup could be similar to those in
Refs.~\cite{Hood:1447,Guthoehrlein:49}.
Note that the concurrence between the 
two cavity fields,
given by Eq.~(\ref{eq:conc_cavities}), is instead
decreasing, and is quickly vanishing. This is due to
the fact that a photon in the cavity is, 
sooner or later, either emitted in the radiated
field, or absorbed or scattered by the cavities mirrors.

\begin{figure}
\includegraphics[width=8.0cm]{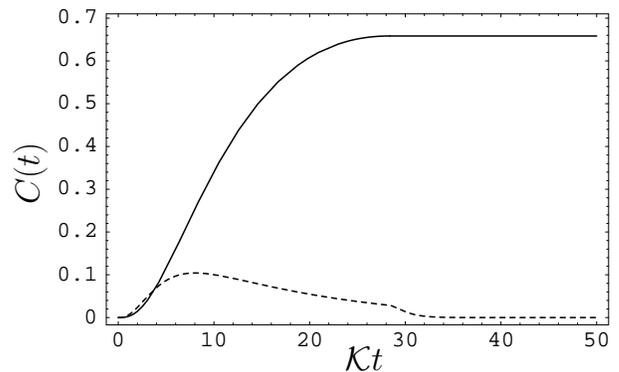}
\caption{The concurrence $C[\rho_{\rm at}(t)]$ between
 the two atoms (solid line) and the concurrence
$C[\rho_{\rm cav}(t)]$
between the two intracavity fields (dashed line) are shown
for the case of equal parameters for the
two subsystems $A$ and $B$.
The parameters are
$g/{\cal{K}} = 10$, $\Omega/{\cal{K}} = 10$, $\Delta/{\cal{K}} = 1000$, and $\kappa/{\cal{K}} = 0.9$.
At $\bar t= 28.32/{\cal{K}} $ the two lasers are switched off,
so that $\Omega= 0$ for $t \geq \bar t$.}
\label{fig:figure_pra_5}
\end{figure}
The behavior for the two concurrences is shown in Fig.~\ref{fig:figure_pra_5}, where we have used
for the two atom-cavity subsystem the same parameters
as in Fig.~\ref{fig:figure_pra_4}, with $\kappa/{\cal{K}}=0.9$.
In this case the two laser beams are turned off
at time $\bar t = 28.32/{\cal{K}}$, where the concurrence 
attains its maximum value of 
$C[\rho_{\rm at}(\bar t)] \simeq 0.66$.
Note that one gets
$\bar t \sim \Delta / (g \Omega)$.
This justifies, in agreement with the discussion in Sec.~\ref{section2}, that
one can omit the effects of spontaneous emissions
in the time interval $[0, \bar t \,]$.
Moreover, also for $t >\bar t$ spontaneous emissions 
are negligible. In fact, when
the two lasers are switched off, the only possibility
to have a jump related to spontaneous
emissions is via the cavity coupling, i.e. 
proportional to the terms $g_a \hat a$ and $g_b \hat b$.
This contribution is negligible not only because
of the large detuning ($\Gamma_a, \Gamma_a',
\Gamma_b, \Gamma_b' \ll 1$),
but it is also vanishing because the cavities are,
for $t \geq \bar t $, practically in the vacuum
state. For example, with the values
used in Fig.~\ref{fig:figure_pra_5}, 
already at $t=\bar t$ the two decaying functions
$\beta(t)$ and $\delta(t)$, cf. Eqs.~({\ref{eq:sol_tbar1})
and (\ref{eq:delta_tbar}), have negligible values,
$|\beta(\bar t)|^2 \simeq |\delta(\bar t)|^2 \simeq 0.01$.

\section{Conditional preparation of entanglement}
\label{section4}

Let us now turn our attention to the case of
a conditional preparation of the entanglement between
the two atoms, and its subsequent storage.
This new situation is obtained by introducing
a photodetector of quantum efficiency $\eta$ 
that monitors the radiated field, as
indicated in Fig.~\ref{fig:figure_pra_1}.
We are interested to study the case when ``no click" 
occurs, i.e. a conditional evolution under imperfect
detection.
When a ``click" at the photodetector is recorded,
the conditional preparation is not successful,
and the preparation procedure has to be repeated again.

In order to properly treat this problem, 
let us introduce the following consideration, 
cf.~\cite{Carmichael:1200} and Appendix~\ref{app:A}.
As it has been already mentioned
in Sec.~\ref{section2},
the probability for a jump 
$\hat J_i$ to occur in the time 
interval $[t, t + dt )$ is given by $p_{\rm i}(t) \!=\! 
\langle \hat J_i^\dagger \hat J_i  \rangle_t \, d t$.
The increment in this time interval
for $p_{\rm yes}(t)$, cf. Eq.~(\ref{eq:p_yes}), is equal to
\begin{equation}
dp_{\rm yes}(t) = \langle \hat J_1^\dagger \hat J_1  \rangle_t \, d t + \langle \hat J_2^\dagger \hat J_2  \rangle_t \, d t +
\langle \hat J_3^\dagger \hat J_3  \rangle_t \, d t \, .
\label{eq:d_p_yes}
\end{equation}
Using Eqs.~(\ref{jump_1}) and (\ref{jumps_2_3})
one obtains, by integrating Eq.~(\ref{eq:d_p_yes}), 
that  
\begin{equation}
p_{\rm yes}(t) = p_{\rm rad}(t) +  p_{\rm abs}(t) \, ,
\label{eq:p_yes_int}
\end{equation}
where
\begin{eqnarray}
p_{\rm rad}(t) \!\!&=&\!\! \!\! \int_0^t \! dt' 
\langle \hat J_1^\dagger \hat J_1  \rangle_{t'}
\!=\! \kappa_a \!\! \int_0^t \! dt'  |\beta(t')|^2 
\!+\! \kappa_b \!\! \int_0^t \!dt'  |\delta (t')|^2 \nonumber \\
\!&+&\! 2 \, \sqrt{\kappa_a \kappa_b} \int_0^t \!
dt' {\rm Re} \! \left[ \beta^{*}(t') \delta (t')e^{-i\phi} \right]\, ,
\label{eq:p_rad_t}
\end{eqnarray}
and 
\begin{eqnarray}
p_{\rm abs}(t) \!&=&\! \int_0^t \! dt' 
\langle \hat J_2^\dagger \hat J_2  \rangle_{t'}
+ \int_0^t \! dt' 
\langle \hat J_3^\dagger \hat J_3  \rangle_{t'} \nonumber \\
\!&=&\! \kappa_a' \int_0^t \! dt'  |\beta(t')|^2 
+ \kappa_b' \int_0^t \! dt'  |\delta (t')|^2 \, .
\label{eq:p_abs_t}
\end{eqnarray}
The function $p_{\rm rad}(t)$ represents the probability
that a photon
is radiated by the cascaded system 
in the time interval $[0, t]$, and $p_{\rm abs}(t)$
the probability that a photon is absorbed or scattered 
by the cavity mirrors in the same time interval.
Note that because $\delta (t)$ contains an
overall factor $e^{i\phi}$, cf. Eqs.~(\ref{eq:diff_eq_sol_general_g_d}) and (\ref{eq:fpm}), the phase $\phi$ is irrelevant in Eq.~(\ref{eq:p_rad_t}).

Let us now assume that
somehow we know that for sure
in the time interval $[0,t]$
a photon has been released by the cascaded system
into its environment.
In this case one would have that 
$p_{\rm yes}(t)\!=\!1$, and $p_{\rm no}(t)\!=\!0$,
i.e. we know for sure in which of the two possible 
realizations the system is found at time $t$.
The density operator $\hat \rho (t)$ that describes the
system is then given, cf. Eqs.~(\ref{eq:rho_yes}) and (\ref{eq:rho_t}),
by  $\hat \rho (t) \!=\! \hat \rho_{\rm yes} (t)\!=\!|e\rangle \langle e|$. It follows that the two atoms are in the 
separable state $|0,0\rangle$, and, obviously, the related concurrence is equal to zero. 
The release of a photon in the environment
destroys any entanglement between the two atoms.

If we now assume that we are in the opposite case,
i.e. that
somehow we know that for sure
in the time interval $[0,t]$
a photon has not been released by the cascaded system
into its environment,
then $p_{\rm yes}(t)\!=\!0$, and $p_{\rm no}(t)\!=\!1$.
The density operator $\hat \rho (t)$ that describes the
system is given, in this case,
by  $\hat \rho (t) \!=\! \hat \rho_{\rm no} (t)$,
cf.~Eq.~(\ref{eq:rho_no}). 
The reduced density operator of the system consisting
of the two atoms, $\hat \rho_{\rm at| no} (t) \!=\! {\rm Tr}_{\rm cav} \left[ \hat \rho_{\rm no} (t) \right]$, is now given by
\begin{eqnarray}
\hat \rho_{\rm at|no} (t) \!\!&=&\!\! 
\frac{1}{p_{\rm no}(t)} 
\Big\{ |\alpha(t)|^2 |1,0 \rangle \langle 1,0| \!+\!
|\gamma(t)|^2 |0,1 \rangle \langle 0,1| \nonumber \\
\!\!&+&\!\! \alpha(t)\gamma^{*}(t)|1,0 \rangle \langle 0,1|
+\alpha^{*}(t)\gamma(t)|0,1 \rangle \langle 1,0|
\nonumber \\
\!\!&+&\!\! \left[|\beta(t)|^2 + |\delta(t)|^2   
\right]   |0,0 \rangle \langle 0,0| \Big\} \, ,
\label{eq:rho_atoms_no}
\end{eqnarray}
where 
$p_{\rm no}(t)$ is given by Eq.~(\ref{eq:p_no}).
It is easy to show that from Eq.~(\ref{eq:conc}) the
concurrence $C[\rho_{\rm at| no}(t)]$ is equal to
\begin{equation}
C[\rho_{\rm at|no}(t)] = \frac{2 \left| \alpha(t) \right| \left| 
\gamma(t)  \right|}{p_{\rm no}(t)} = 
\frac{C[\rho_{\rm at}(t)]}{p_{\rm no}(t)}\, ,
\label{eq:conc_atoms_no}
\end{equation}
where we have also used Eq.~(\ref{eq:conc_atoms}).
Because $p_{\rm no}(t)\!\leq\! 1$, it follows that
$C[\rho_{\rm at|no}(t)] \!\geq\! C[\rho_{\rm at}(t)]$. 
In other words, the knowledge that no photon has been 
released by the cascade system increases the entanglement
between the two atoms.

Let us now consider the case 
when a photodetector of given efficiency $\eta$ is
used to monitor the radiated field.
It is possible to show, by using the quantum trajectory 
method~\cite{Carmichael:1200,Carmichael:private},
see Eq.~(\ref{eq:p_0_1}) in the Appendix~\ref{app:A},
that the probability of not recording a click at the 
photodetector up to time $t$ is given by
\begin{eqnarray}
p_{0} (t) \!&=&\! p_{\rm no} (t) + (1-\eta) p_{\rm rad}(t)+ p_{\rm abs} (t) 
= 1 - \eta p_{\rm rad}(t) \nonumber \\
\!&=&\! 1 - \eta + \eta \left[
p_{\rm no} (t) + p_{\rm abs} (t) \right] \, ,
\label{eq:p_no_D}
\end{eqnarray}
where we have also used Eqs.~(\ref{eq:p_yes_int})
and ~(\ref{eq:p_yes}).
The conditional state given that the detector does not record
a photon, is a weighted sum over the conditional
density operators reached via the various records of
this type, i.e. null-measurement at the detector that
monitors the radiated field. This yields, 
cf.~Eq.~(\ref{eq:rho_0_1}),
\begin{eqnarray}
 \hspace*{-0.0cm} \hat \rho_{0} (t) \!&=&\! \frac{1}{p_{0} (t)}
\Big\{ p_{\rm no}(t)\hat \rho_{\rm no}(t) \!+\!
\big[ (1\!-\!\eta) p_{\rm rad}(t) \nonumber \\
&+& p_{\rm abs} (t) \big] \hat \rho_{\rm yes}(t) \Big\}
=\frac{1}{p_{0} (t)} \Big\{
| \bar{\psi}_{\rm no} (t) \rangle \langle
\bar{\psi}_{\rm no} (t) | \nonumber \\
&+&
\big[ (1\!-\!\eta) p_{\rm rad}(t)
\!+\! p_{\rm abs} (t) \big] |e\rangle \langle e| \Big\} \, ,
\label{eq:rho_no_D}
\end{eqnarray}
where $\hat \rho_{\rm no}(t)$ and $\hat \rho_{\rm yes}(t)$
are given by Eq.~(\ref{eq:rho_no}) and Eq.~(\ref{eq:rho_yes}), respectively. Note that if $\eta = 0$
we obtain, from Eq.~(\ref{eq:p_no_D}), that $p_{0}(t)=1$. 
In this case we return to an unconditional evolution,
and $\hat \rho_{0} (t)$
becomes again $\hat \rho (t)$, cf. Eq.~(\ref{eq:rho_no_D})
and Eq.~(\ref{eq:rho_t_1}). 
For $\eta = 1$, and no photon absorption
or scattering by the cavity mirrors, i.e. when 
$p_{\rm abs}(t)=0$, 
we obtain, from Eq.~(\ref{eq:p_no_D}), that $p_{0}(t)=p_{\rm no}$. 
This is the case where we can be sure that no photon
has been lost by the system,
and $\hat \rho_{0} (t)$
becomes $\hat \rho_{\rm no} (t)$, cf. Eq.~(\ref{eq:rho_no_D})
and Eq.~(\ref{eq:rho_no}).

For the reduced density operator of the system consisting
of the two atoms we have that
\begin{eqnarray}
\hspace*{-0.5cm}\hat \rho_{{\rm at}| 0} (t) \!\!&=&\!\! {\rm Tr}_{\rm cav} \left[ \hat \rho_{0} (t) \right] 
= \frac{p_{\rm no}(t)}{p_0(t)}
\hat \rho_{\rm at|no} (t) \nonumber \\ 
&+& \frac{\big[ (1\!-\!\eta) p_{\rm rad}(t)
\!+\! p_{\rm abs} (t) \big]}{p_{0}(t)} |0,0 \rangle 
\langle 0,0| \, ,
\label{eq:rho_atoms_no_D}
\end{eqnarray}
where $\hat \rho_{\rm at|no} (t)$ is given 
by Eq.~(\ref{eq:rho_atoms_no}).
Considering now the $4\times4$ density matrix 
$\rho_{{\rm at}| 0}(t)$, related to the density operator of Eq.~(\ref{eq:rho_atoms_no_D}) in the basis 
$\{ |0,0 \rangle, |0,1 \rangle, |1,0 \rangle,
|1,1 \rangle \}$, it is easy to show that the
concurrence $C[\rho_{{\rm at}|0}(t)]$ is, using 
Eq.~(\ref{eq:conc}), given by
\begin{equation}
C[\rho_{{\rm at}|0}(t)]
= \frac{1}{p_0(t)} \, C[\rho_{\rm at}(t)]\, ,
\label{eq:conc_atoms_no_D}
\end{equation}
where we have used Eq.~(\ref{eq:conc_atoms_no}).
This is the concurrence between the two atoms
in the presence of a photodetector, of
given efficiency, informing us that 
no photon has been registered
in the radiated field.
If $\eta = 0$, we have $p_0(t)=1$,
and 
the concurrence is again equal to $C[\rho_{\rm at}(t)]$,  
as in Eq.~(\ref{eq:conc_atoms}).
Note that $p_{0}(t)\leq 1$,
so that the concurrence 
$C[\rho_{{\rm at}|0}(t)] \geq  C[\rho_{\rm at}(t)]$.
Moreover, $p_0(t)$ is a decreasing function with $\eta$, so that for $\eta =1$ it reaches 
its minimum value, and, consequently,
cf.~Eq.~(\ref{eq:conc_atoms_no_D}), the concurrence
$C[\rho_{{\rm at}|0}(t)]$ reaches its maximum value.
In this case,
and for perfect mirrors, i.e. for $p_{\rm abs}(t) \!=\! 0$, 
one has $p_0(t)\!=\! p_{\rm no}(t)$, and the concurrence 
is given by Eq.~(\ref{eq:conc_atoms_no}).

Let us now consider the case analyzed
in Fig.~\ref{fig:figure_pra_5}, but with the
presence of a
detector of efficiency $\eta$ that monitors
the radiated field.
Because 
$|\beta(t)|^2 \simeq |\delta(t)|^2\simeq 0$,
for ${\cal{K}} t \gg 1$,
this implies that in this case one has, 
using Eq.~(\ref{eq:p_no}), 
$p_{\rm no}(t) \!\simeq\! |\alpha(\bar t)|^2 \!+\! |\gamma(\bar t)|^2$. With the values
considered in Fig.~\ref{fig:figure_pra_5},
this gives $p_{\rm no}(t) \simeq 0.66$. 
Moreover, from Eq.~(\ref{eq:p_abs_t}), and
using the same parameters 
as in Fig.~\ref{fig:figure_pra_5}, we
obtain, for ${\cal{K}} t \gg 1$,
the value  $p_{\rm abs}(t) \simeq 0.2$ .
Considering that the single-photon
detector-efficiency has already reached a value 
of approximately $\eta = 0.88$, cf. Ref.~\cite{Yamamoto:1063},
from Eq.~(\ref{eq:p_no_D}) we obtain, for these parameters, the value 
$p_0(t)\simeq 0.88$.
Because for $t > \bar t$ the concurrence 
$C[\rho_{\rm at}(t)]$ is 
given by Eq.~(\ref{eq:conc_atoms_tbar}), i.e. 
$C[\rho_{\rm at}(t)] \simeq 0.66$, we obtain from 
Eq.~(\ref{eq:conc_atoms_no_D}) that, 
for ${\cal{K}} t \gg 1$, 
$C[\rho_{{\rm at}|0}(t)] \simeq 0.75$.
This is the value of the concurrence stored
between the two atoms, with an enhancement of
approximately $14\%$.
Note that in the ideal case of $\eta=1$ and no photon
absorption or scattering by the cavity mirrors,
then $p_0(t) = p_{\rm no}(t)\simeq 0.66$, 
cf. Eq.~(\ref{eq:p_no_D}), so that one would obtain, 
with the chosen parameters,
$C[\rho_{{\rm at}|0}(t)] = C[\rho_{{\rm at}|{\rm no}}(t)] 
\simeq 1$, cf.~Eqs.~(\ref{eq:conc_atoms_no})
and (\ref{eq:conc_atoms_no_D}). This value
of the concurrence is related
to the fact that $|\alpha(\bar t)|\simeq |\gamma(\bar t)|$. 

Finally, an important question is related to the probability of successfully realizing the whole process of
the conditional preparation and storage of entanglement. 
We know that if the detector registers a
photon, then the density operator for the cascaded system
is given by Eq.~(\ref{eq:rho_yes}), 
and no entanglement is present between the two atoms.
The probability of a successful realization of this scheme
is given by the probability that the detector does
not register any photon, probability given by
$p_{0}(t)= 1 - \eta p_{\rm rad}(t)$.
If $\eta \!=\! 0$, i.e. when we are in the case
where no detector is present, this procedure is
always successful, but the concurrence is not enhanced 
and remains, as in the previous section, given by Eq.~(\ref{eq:conc_atoms}).
Note that, because we are interested here in a 
null-measurement
conditional preparation, the above considerations 
remain valid also when a click at the photodetector
is coming from a possible dark count. This
only increases the probability that the whole
procedure has to be repeated from the beginning. 
Using the same parameters as in Fig.~\ref{fig:figure_pra_5}
and a detector efficiency $\eta = 0.88$,
cf. Ref.~\cite{Yamamoto:1063}, we have
that $p_0(t)\simeq 0.88$.
This means that for approximately $88 \%$ of the cases, 
the conditional preparation and storage of the entanglement
between the two atoms, with a value
$C[\rho_{{\rm at}|0}(t)] \simeq 0.75$,
is successfully realized. 
For the remaining cases we have to repeat the whole 
procedure from the beginning. 

\section{conclusions}
\label{conclusion}

The dynamics of a cascaded system that
consists of two atom-cavity subsystems has
been analyzed. 
Considering the two atom-cavity subsystems
driven by a Raman interaction, the evolution of
the open quantum system under study has been described 
by means of a master equation. 
By using the quantum trajectory method,
analytical solutions for the dynamics of the system
have been obtained. 
The entanglement evolution between two stable
ground states for the two atoms,
constituting a two-qubit system,
has been studied using the concurrence.
A similar analysis has been performed for the 
two intracavity fields.

The dynamical evolution of the system shows that 
the two initially disentangled qubits reach states
of significant entanglement.
Moreover, it has been shown that the
entanglement generated between the two atoms can 
be stored by switching off the Raman coupling.
Subsequently, we have analyzed how the entanglement between 
the two atoms can be enhanced,
by monitoring the radiated field with a photodetector of given efficiency, 
via a 
null-measurement conditional preparation.

\section*{ACKNOWLEDGMENTS}
This work was supported by the Deutsche Forschungsgemeinschaft.
The authors thank Howard Carmichael and Adam Miranowicz for 
helpful discussions.

\appendix*
\vspace*{2ex}
\section{}
\label{app:A}

Let us analyze the problem of
the conditional evolution under imperfect 
detection~\cite{Carmichael:1200,Carmichael:private}
by considering a system with two output channels.
Channel~$1$ is monitored by a detector of efficiency 
$\eta$ and
associated with a jump operator $\hat {\cal J}_1$. 
Channel~$2$ is monitored by a detector of unit efficiency,
which could represent the environment into which the system
releases a photon, associated with a jump 
operator $\hat {\cal J}_2$.
One can treat non-unit detection efficiency 
by introducing a beam splitter into channel $1$, 
with transmittivity $\sqrt{\eta}$.
Now the system has three output channels,
the transmitted and the reflected parts
$1_{\rm T}$ and $1_{\rm R}$, respectively,
due to the beam splitter, and 
the original channel 2. 
In principle, all three channels could be thought 
of being monitored by detectors of unit
efficiency. In the following we will indicate
the three channels with $i= 1_{\rm T}, 1_{\rm R}, 2$.
We are interested in the conditional evolution
under null-measurement at the photodetector 
that monitors the transmitted beam, i.e. channel $1_{\rm T}$. 
The master equation for the density operator
that describes the system can be formally 
written, cf.~\cite{Carmichael2}, as
$d\hat \rho(t)/dt = {\cal L} \hat 
\rho(t)$, 
where the superoperator
${\cal L} = {\cal L}_B + {\cal S}_{1_{\rm T}}
+ {\cal S}_{1_{\rm R}} +  {\cal S}_2$
is in the 
usual Lindblad form.
The between-jump superoperator ${\cal L}_B$ 
is given by 
\begin{equation}
{\cal L}_B \, \cdot = \frac{1}{i\hbar} \left[\hat H_S, \cdot \right] 
- \frac{1}{2} \sum_i \left( \hat {\cal J}_i^\dagger \hat 
{\cal J}_i \cdot
+ \cdot\hat {\cal J}_i^\dagger \hat {\cal J}_i \right)\, ,
\label{eq:BJsup_operator}
\end{equation}
where $\hat H_S$ is the system Hamiltonian,
with the jump operators for the transmitted and
the reflected channels given by
\begin{equation}
\hat {\cal J}_{1_{\rm T}} = \sqrt{\eta} \, \hat {\cal J}_1 \, , ~~
\hat {\cal J}_{1_{\rm R}} = \sqrt{1- \eta} \, \hat {\cal J}_1 \, .
\end{equation}
The three jump superoperators are defined as
\begin{eqnarray}
{\cal S}_i \, \cdot = \hat {\cal J}_i  \, \cdot \, \hat {\cal J}_i^\dagger 
~~~~(i= 1_{\rm T}, 1_{\rm R}, 2) \, .
\label{eq:Jsup_operators_1}
\end{eqnarray}

Let the initial density operator be $\hat \rho(0)$. 
Because we are interested in a system where we can have 
at most one jump, there are four records of interest.
First, the record where neither detector clicks in the 
interval $[0, t)$; for it we have the
probability and conditional density operator~\cite{Carmichael2}
\begin{equation}
\hspace*{-0.2cm} p_{\rm no}(t) = {\rm Tr} \! \left[e^{{\cal L}_B t}  \hat \rho(0) \right] , ~\hat \rho_{\rm no}(t) = \frac{e^{{\cal L}_B t}  
\hat \rho(0)}{{\rm Tr} \! \left[e^{{\cal L}_B t}  \hat \rho(0) \right]} \, .
\label{eq:p_00}
\end{equation}
We have then the record given by a photon detected 
in the time interval $[t', t'+dt)$, with $t' < t$,
at one of the detectors $i$ ($i = 1_{\rm T},1_{\rm R},2$)
and the other two detectors not clicking. 
For it we have the probability
and conditional density operator ($i = 1_{\rm T},1_{\rm R},2$)
\begin{eqnarray}
& &p_i(t) = {\rm Tr} \! 
\left[ e^{{\cal L}_B (t-t')} {\cal S}_i e^{{\cal L}_B t'}  \hat \rho(0) \right] dt \, , \nonumber \\
& &\hat \rho_i(t) = \frac{e^{{\cal L}_B (t-t')} {\cal S}_i e^{{\cal L}_B t'}  \hat \rho(0)}{{\rm Tr} \! \left[e^{{\cal L}_B (t-t')}{\cal S}_i e^{{\cal L}_B t'}  \hat \rho(0) \right]} \, .
\label{eq:p_t0}
\end{eqnarray}
The probability for no click at detector $1_{\rm T}$
up to time $t$ is given by a sum over all events
with no click in channel $1_{\rm T}$,
\begin{equation}
p_0(t) = p_{\rm no}(t) + \int_0^t dt' p_{1_{\rm R}}(t') +
\int_0^t dt' p_2(t') \, .
\label{eq:p_0}
\end{equation}
The corresponding conditional density operator is a weighted sum over conditional density operators of the form
\begin{eqnarray}
\hat \rho_0(t) \!&=&\! \frac{1}{p_0(t)} 
\Big[ p_{\rm no}(t) \hat \rho_{\rm no}(t) + \int_0^t dt' 
p_{1_{\rm R}}(t') \hat \rho_{1_{\rm R}}(t')  \nonumber \\
\!&+& \int_0^t dt' 
p_2(t') \hat \rho_2(t')   \Big] \, .
\label{eq:rho_0}
\end{eqnarray}
Following the quantum trajectory 
method~\cite{Carmichael2}, when no jump occurs,
the system evolves between time $t_0=0$ and $t$
via
\begin{equation}
\hat \rho_{\rm no}'(t) = e^{{\cal L}_B t} \hat \rho(0) \, ,
\label{eq:schr_non_unitary_app}
\end{equation}
where $\hat \rho_{\rm no}'(t)$ is, in general, not normalized. 
The evolution governed by Eq.~(\ref{eq:schr_non_unitary_app})
is randomly interrupted by jumps.
If a jump occurres at time
$t_{\rm J}$, $t_{\rm J} \in [0, t)$, the
density operator collapses into $|0 \rangle\langle 0|$, 
\begin{equation}
\hspace*{0cm} {\cal S}_i \hat \rho_{\rm no}'(t_{\rm J}) =
\hat {\cal J}_{i} \hat \rho_{\rm no}'(t_{\rm J})
\hat {\cal J}_{i}^\dagger \rightarrow |0 \rangle \langle 0| 
~~~(i= 1_{\rm T}, 1_{\rm R}, 2). 
\label{eq:jump_op_i_app} 
\end{equation}
Let us now define 
\begin{eqnarray}
\hspace*{-0.51cm}
P_1(t) &=& \int_0^t \!\!\! dt' \!
\langle \hat {\cal J}_{1_{\rm T}}^\dagger \hat {\cal J}_{1_{\rm T}}\rangle_{t'} + \int_0^t \!\!\! dt' \!
\langle \hat {\cal J}_{1_{\rm R}}^\dagger 
\hat {\cal J}_{1_{\rm R}}\rangle_{t'} \, , \nonumber \\
P_2(t) &=& \int_0^t \!\!\! dt' \!
\langle \hat {\cal J}_{2}^\dagger \hat {\cal J}_{2}\rangle_{t'} \, ,
\label{eq:p_i_t_app}
\end{eqnarray}
where $\langle \ldots \rangle_{t'}= {\rm Tr}[ \hat \rho_{\rm no}'(t')\ldots]$\,. The function $P_1(t)$ represents the probability
that in the time interval $[0, t)$ a photon
is emitted by the system into channel 
$1_{\rm T}$ or $1_{\rm R}$, and $P_2(t)$ is
the probability that a photon is emitted into channel 2.
The probability that the detector $1_{\rm T}$ 
does not click up to time $t$ is obtained 
from Eq.~(\ref{eq:p_0}) as
\begin{eqnarray}
\hspace*{-0.4cm}& & p_0(t) = 
p_{\rm no}(t) \!+\!(1 \!-\! \eta) P_1(t) \!+\! P_2(t) \, ,
\label{eq:p_0_1}
\end{eqnarray}
with $p_{\rm no}(t)$ given in Eq.~(\ref{eq:p_00}).
Finally, the conditional density operator given that the detector 
$1_{\rm T}$ does 
not click is 
given, using Eq.~(\ref{eq:rho_0}), by
\begin{eqnarray}
\hspace*{-0.5cm}
\hat \rho_0(t) \!=\! \frac{1}{p_0(t)} 
\Big\{ \hat \rho_{\rm no}'(t) \!+\! 
\big[ (1 \!-\! \eta)P_1(t) \!+\! P_2(t)  \big] |0 \rangle \langle 0| \Big\} .
\label{eq:rho_0_1}
\end{eqnarray}

\end{document}